\documentclass[prb,twocolumn]{revtex4}
\usepackage{epsf}
\usepackage{times}
\begin{document}

\title{Semiclassical theory of spin-polarized shot noise in 
mesoscopic diffusive conductors}

\author{M. Zareyan$^{1,2}$ and W. Belzig $^{3}$}

\affiliation{
$^{1}$ Max-Planck-Institute f\"ur Physik komplexer Systeme,
N\"othnitzer Str. 38, 01187 Dresden, Germany\\
$^{2}$ Institute for Advanced Studies in Basic Sciences, 45195-159,
Zanjan, Iran\\
$^{3}$ Departement f\"ur Physik und Astronomie, Klingelbergstr. 82,
4056 Basel, Switzerland}

\date{\today}
\begin{abstract}
  We study fluctuations of spin-polarized currents in a three-terminal
  spin-valve system consisting of a diffusive normal metal wire
  connected by tunnel junctions to three ferromagnetic terminals.  Based
  on a spin-dependent Boltzmann-Langevin equation, we develop a
  semiclassical theory of charge and spin currents and the correlations
  of the currents fluctuations. In the three terminal system, we show
  that current fluctuations are strongly affected by the spin-flip
  scattering in the normal metal and the spin polarizations
  of the terminals, which may point in different directions. 
  We analyze the dependence of the shot noise and the cross-correlations
  on the spin-flip scattering rate in the full range of the spin
  polarizations and for different magnetic configurations. Our result
  demonstrate that noise measurements in multi-terminal devices allow to
  determine the spin-flip scattering rate by changing the polarizations
  of ferromagnetic terminals. 
\end{abstract} 

\maketitle

\section{Introduction}

Spin-dependent electronic transport and use of the spin degree of
freedom of electrons in hybrid magnetic structures have recently been
subject of highly interesting field termed spintronics. This developing
field has emerged from many exciting phenomena, the giant
magneto-resistance being the most well known example, which has made it
attractive for applications as well as fundamental studies
\cite{meservey94,levy94,gijs97,prinz98,johnson:85,Jedema00,zaffalon:03,awschalom02}.
Typical studied systems are based on the use of ferromagnetic metals (F)
and/or external magnetic fields to inject, manipulate and detect
spin-polarized electrons inside a mesoscopic normal metal (N).
\par
An important characteristic of mesoscopic systems 
is the appearance of shot noise, the fluctuations of current 
through the system due to the randomness of the electronic 
scattering processes and the quantum statistics. Shot 
noise and nonlocal correlations of the current fluctuations 
contain additional information on the conduction process 
which is not gainable through 
a mean current measurement. In ferromagnetic-normal metal 
structures, in which the spin of electrons plays an essential rule, 
current fluctuations are due to the randomness of both charge 
and spin transport processes. Thus, shot noise measurements are
expected to provide information about spin-dependent scattering 
processes and spin accumulation in the system. Together 
with the importance of noise in spintronic devices in view of 
applications this motivates our study of spin-polarized shot 
noise.  
\par
In the past years, shot noise has been extensively studied in a wide
variety of hybrid structures involving normal metals, semiconductors and
superconductors\cite{blanterRev,nazarov:03}. However, there are few
studies devoted to shot noise in ferromagnet-normal metal systems.
Results of earlier studies of current fluctuations in FNF double
barrier systems in the Coulomb blockade regime\cite{bulka99} and FIF (I
being an insulator) systems \cite{nowak99} can be understood in terms of
well known results for the corresponding normal-metal systems for two
spin directions\cite{blanterRev}.  Tserkovnyak and Brataas studied shot
noise in double barrier FNF systems with noncollinear magnetizations in
F-terminals \cite{tserkovnyak01}. They found that the shot noise has a
non-monotonic behavior with respect to the relative angle of the
magnetizations for different scattering regimes and different types of
FN junctions.
\par
The effect of spin flip scattering on spin-polarized current 
fluctuations has been considered in Refs. 
\onlinecite{mishchenko03,sanchez03,mishchenko04,lamacraft04,BZ04}.
Mishchenko \cite{mishchenko03} found that in a perfectly polarized 
two-terminal double barrier system spin-flip scattering leads to 
a strong dependence of shot noise on the relative orientation of 
the magnetizations in F-terminals.  In Ref. \onlinecite{BZ04} 
we have proposed a four-terminal spin-valve 
system of tunnel junctions to study spin-dependent shot noise and cross-
correlations simultaneously. It has been found that 
the cross-correlations between currents in terminals 
with opposite spin polarization can be used to measure 
directly the spin-flip scattering rate.
\par
Recently, there have been also studies of the current 
fluctuations of spin-polarized entangled electrons in 
quantum dots  and wires \cite{egues02} and in quantum 
dots attached to the ferromagnetic leads in the Coulomb 
blockade \cite{lu02}, Kondo \cite{sanchez04} and sequential 
tunneling \cite{bulka00,cottet04,cottet04a,cottet04b,sauret:04} regimes. 
\par
In this paper, we study current fluctuations in a three terminal
diffusive FNF system in the full range of spin polarizations and the
spin-flip scattering intensity. Based on the Boltzmann-Langevin
\cite{kadomtsev57,kogan69,nagaev92,sukhorukov99} kinetic approach, we
develop a semiclassical theory for spin-polarized transport in the
presence of the spin-flip scattering. We obtain the basic equations of
charge and spin transport, which allow the calculation of mean currents
and the correlations of current fluctuations in multi-terminal
diffusive systems. Applying the developed formalism to a
three-terminal geometry, we find that current correlations are affected
strongly by spin-flip scattering and spin polarizations.  We
focus on the shot noise of the total current through the system and the
cross-correlations measured between currents of two terminals. We
demonstrate how these correlations deviate from the noise
characteristics of the unpolarized system, depending on the spin-flip
scattering rate, the polarizations of the terminals and their magnetic
configurations (relative directions).  Our results provide a full
analysis for spin-dependent shot noise and cross-correlations in terms
of the relevant parameters.
\par
The structure of the paper is as follows.  In section II we extend the
Boltzmann-Langevin equations to the diffusive systems, in which
spin-flip scattering takes place and which are connected to
ferromagnetic terminals.  We find the basic equations of the charge and
the spin currents and the correlations of their fluctuations.  In
section III we apply this formalism to the three terminal system. We
obtain the Fano factor and the cross-correlations between currents
through two different terminals.  Section VI is devoted to the analysis
of the calculated quantities.  We present analytical expressions for the
Fano factor and the cross-correlations in different important limits.
Finally, we end with some conclusion in section V.

\section{Boltzmann-Langevin equations with spin-flip 
scattering}

In this section we extend the semiclassical Boltzmann-Langevin kinetic
approach \cite{blanterRev} to cover spin-polarized transport. In the
presence of spin-flip scattering the Boltzmann-Langevin equation is
written as
\begin{eqnarray}
\frac{d}{dt} f_{\alpha}=I^{{\text {imp}}}[f_{\alpha}]+
I_{\alpha}^{{\text {sf}}}[f_{\alpha},f_{-\alpha}] +\xi^{{\text
{imp}}}_{\alpha}+\xi^{{\text {sf}}}_{\alpha}, \label{BLs}
\end{eqnarray}
where the  fluctuating distribution function of spin
$\alpha$($=\pm 1$) electrons 
$
f_{\alpha}({\bf p},{\bf r},t)={\bar  f}_{\alpha}
({\bf  p},{\bf r}) +\delta f_{\alpha}({\bf p},
{\bf r},t)
$
depends  on the momentum ${\bf p}$, the  position ${\bf r}$,
and the time $t$. Eq.~(\ref{BLs}) contains both normal 
impurity and spin-flip collision integrals which are given 
by the relations 
\begin{eqnarray}
&&
I^{{\text {imp}}}[f_{\alpha}]=\Omega \int \frac{d{\bf p}^{\prime}}
{(2\pi h)^3} [J_{\alpha \alpha}({\bf p}^{\prime},{\bf p})- 
J_{\alpha \alpha}({\bf p},{\bf p}^{\prime})],
 \label{Is}
\\*
&&
I_{\alpha}^{{\text {sf}}}[f_{\alpha},f_{-\alpha}]= 
\Omega \int \frac{d{\bf p}^{\prime}}
{(2\pi h)^3} [J_{-\alpha \alpha}({\bf p}^{\prime},{\bf p})-
J_{\alpha -\alpha}({\bf p},{\bf p}^{\prime})].
\nonumber \\*
&& 
\label{Issf}
\end{eqnarray}
Here
$
J_{\alpha \alpha ^{\prime}}({\bf p},{\bf p}^{\prime},{\bf
r},t)= W_{\alpha \alpha ^{\prime}}({\bf p},{\bf p}^{\prime},{\bf
r}) f_{\alpha}({\bf   p},{\bf r},t)[1-f_{\alpha ^{\prime}}({\bf
p}^{\prime} ,{\bf r},t)], 
$ where 
$
W_{\alpha \alpha^{\prime}}({\bf p},{\bf p}^{\prime},{\bf r})
$ 
is the elastic scattering rate from the state ${\bf p}, \alpha$
into ${\bf p}^{\prime}, \alpha ^{\prime}$; $\Omega$ is the volume 
of the system. The corresponding Langevin sources of fluctuations 
due to  the random character of the  electron scattering are given 
by
\begin{eqnarray}
&&
\xi^{{\text {imp}}}_{\alpha}=\Omega \int \frac{d{\bf p}^{\prime}}
{(2\pi h)^3} [\delta J_{\alpha \alpha}({\bf p}^{\prime},{\bf p})
-\delta J_{\alpha \alpha}({\bf p},{\bf p}^{\prime})], 
\label{ksis}
\\*
&&
\xi^{\text {sf}}_{\alpha}=\Omega \int \frac{d{\bf p}^{\prime}}{(2\pi
h)^3} [\delta J_{-\alpha \alpha}({\bf p}^{\prime},{\bf p})- \delta
J_{\alpha -\alpha}({\bf p},{\bf p}^{\prime})], 
\label{ksisfs}
\end{eqnarray}
where the random variable 
$
\delta J_{\alpha \alpha^{\prime}}({\bf p}^{\prime}
,{\bf p},{\bf r},t) 
$
is the fluctuation of the current 
$
J_{\alpha \alpha ^{\prime}}({\bf p},{\bf p}^{\prime}, {\bf
r},t)= {\bar J}_{\alpha \alpha^{\prime}}({\bf p}, {\bf
p}^{\prime},{\bf r},t)+ \delta J_{\alpha \alpha^{\prime}}({\bf
p},{\bf p}^{\prime},{\bf r},t);
$ 
with
$
{\bar J}_{\alpha \alpha ^{\prime}}({\bf p}, {\bf
p}^{\prime},{\bf r},t)= W_{\alpha \alpha ^{\prime}} ({\bf p},{\bf
p}^{\prime},{\bf r}) {\bar f}_{\alpha}({\bf p} ,{\bf r},t)[1-{\bar
f}_{\alpha ^{\prime}}({\bf p}^{\prime}, {\bf r},t)] 
$
being the mean current.
\par
We will assume that all scattering events are independent elementary
processes and thus the correlator of the current fluctuations
$\delta J_{\alpha \alpha ^{\prime}}({\bf p},{\bf p}^{\prime},{\bf
r},t)$ is that of a Poissonian process:
\begin{eqnarray}
&& <\delta J_{\alpha_1 \alpha_2}({\bf p}_1,{\bf p}_2, {\bf r},t)
\delta J_{\alpha_3 \alpha_4 }({\bf p}_3 ,{\bf p}_4,{\bf
r}^{\prime},t^{\prime})>= 
\nonumber \\* 
&&
\frac{(2\pi h)^6}{\Omega}\delta _{\alpha_1\alpha_3} 
\delta_{\alpha_2\alpha_4} \delta({\bf p}_1-{\bf p}_3) 
\delta({\bf p}_2-{\bf p}_4) 
\nonumber \\* 
&&
\times
\delta({\bf r}-{\bf r}^{\prime}) \delta(t-t^{\prime}) 
{\bar J_{\alpha_1 \alpha _2}({\bf p}_1,{\bf p}_2,{\bf
r},t)}.
\label{poisson}
\end{eqnarray}
\par
Due to the non-vanishing spin-flip collision integral in 
(\ref{BLs}), the equations for the distributions of electrons with 
opposite spin directions are coupled. The coupled equations can be 
transformed into two decoupled equations for the charge 
$f_{\text {c}}=\sum _{\alpha} f_{\alpha}/2$ 
and spin 
$f_{\text {s}}=\sum _{\alpha} \alpha f_{\alpha}/2$
distribution functions, which read
\begin{eqnarray}
\frac{d}{dt} f_{{\text {c}}(\text {s})}=I^{{\text {imp}}}
[f_{{\text {c}}({\text {s}})}]
+I^{{\text {sf}}}_{{\text {c}}({\text {s}})}
[f_{{\text {c}}{(\text {s})}}]+ \xi^{{\text {imp}}}
_{{\text {c}}({\text {s}})}+
\xi^{{\text {sf}}}_{{\text {c}}({\text {s}})}.
\label{BLfp}
\end{eqnarray}
Here we have introduced different collision integrals: 
\begin{eqnarray}
&&
I^{{\text {imp}}}[f_{{\text {c}}({\text {s}})}]=\Omega\int \frac{d{\bf
p}^{\prime}} {(2\pi h)^3}[J^{{\text {imp}}}_{{\text {c}}({\text {s}})}({\bf
p}^{\prime},{\bf p})- J^{{\text {imp}}}_{{\text {c}}({\text {s}})}({\bf
p},{\bf p}^{\prime})],\nonumber \\*
&&
\label{If}\\*
&&
I_{\text {c}}^{{\text {sf}}}[f_{\text {c}}]
=\Omega\int \frac{d{\bf p}^{\prime}}
{(2\pi h)^3}[J_{\text {c}}^{{\text {sf}}}({\bf p}^{\prime},{\bf p})-
J_{\text {c}}^{{\text {sf}}}({\bf p},{\bf p}^{\prime})],
\label{Isff}\\*
&&
I^{{\text {sf}}}_{\text {s}}[f_{s}]=-\Omega\int \frac{d{\bf
p}^{\prime}} {(2\pi h)^3}W^{{\text {sf}}}({\bf p},{\bf
p}^{\prime}) [f_{\text {s}}({\bf p}^{\prime})+f_{\text {s}} ({\bf
p})],\nonumber \\*
&&
\label{Isffp}
\end{eqnarray}
where
\begin{eqnarray}
&&
J^{{\text {imp}}({\text {sf}})}_{{\text {c}}({\text {s}})}({\bf p},{\bf
p}^{\prime},{\bf r},t)= W^{{\text {imp}}({\text {sf}})}({\bf p},{\bf
p}^{\prime},{\bf r})
\nonumber \\*
&&
\times f_{{\text {c}}({\text {s}})}({\bf p},{\bf r},t)
[1-f_{{\text {c}}({\text {s}})}({\bf p}^{\prime} ,{\bf r},t)],
\end{eqnarray}
and we assumed 
$ W_{\alpha\alpha^{\prime}} ({\bf  p},{\bf
p}^{\prime}) =W_{\alpha ^{\prime}  \alpha} ({\bf p}^{\prime},{\bf
p})$, $W_{+-}=W_{-+}=W^{{\text{sf}}}; $
$
W_{++}=W_{--}=W^{{\text{imp}}}.$
The corresponding  Langevin sources of fluctuations are given by
\begin{eqnarray}
&&
\xi_{{\text {c}}}^{{\text {imp}}({\text {sf}})}({\bf p},{\bf r},t)
=\frac{1}{2}
\sum_{\alpha} \xi^{{\text {imp}}({\text {sf}})}_{\alpha}({\bf
p},{\bf r},t), \label{ksiimsf}\\*
&&
 \xi^{{\text{imp}}({\text {sf}})}_{\text {s}}({\bf p},{\bf r},t)
=\frac{1}{2}\sum_{\alpha} \alpha 
\xi^{{\text {imp}}({\text {sf}})}_{\alpha}({\bf p},{\bf r},t).
\label{ksiimsfs}
\end{eqnarray}
\par 
In the following we assume that all the quantities are sharply 
peaked around the Fermi energy and instead of ${\bf p}$ use the 
quantities $\varepsilon$ the energy and ${\bf n}$ the direction of 
the Fermi momentum. Then, for elastic scattering of electrons the 
following relation holds
\begin{eqnarray}
\Omega W^{{\text {imp}}({\text {sf}})}({\bf p},{\bf p}^{\prime}
,{\bf r})=\frac{2}{N_0}\delta (\varepsilon-\varepsilon^{\prime})
w^{{\text {imp}}({\text {sf}})} ({\bf n},{\bf n}
^{\prime},{\bf r}),
\label{wnnp}
\end{eqnarray}
where $N_{0}$ is the density of states in the Fermi level.
\par
For a diffusive conductor we apply the standard diffusive
approximation to the kinetic equations (\ref{BLfp})
where the charge and spin distribution functions are split into
the symmetric and asymmetric parts:
\begin{eqnarray}
f_{{\text {c}}({\text {s}})}({\bf n},\varepsilon,{\bf r},t)= 
f_{{\text {c}}({\text{s}})0}(\varepsilon,{\bf r},t) 
+{\bf n}.{\bf f}_{{\text {c}}({\text {s}})1}(\varepsilon,{\bf r},t).
\label{fnex}
\end{eqnarray}
Substituting this form of $f_{\text {c}}$ in Eqs. (\ref{BLfp}) and averaging 
subsequently over the Fermi momentum direction first weighted with 
one and then with ${\bf n}$, we obtain 
\begin{eqnarray}
&&
\frac{v_{F}}{3} \nabla . {\bf f}_{{\text {c}}1}= 
\int d{\bf n} \xi_{\text {c}}^{{\text {sf}}}({\bf n},\varepsilon,{\bf r},t),
\label{df}\\*
&&
\frac{v_{F}}{3}\nabla f_{{\text {c}}0}=
-\frac{1}{3\tau}{\bf f}_{{\text {c}}1}
\nonumber \\*
&&+\int {\bf
n}d{\bf n} [\xi_{\text {c}}^{{\text {imp}}}({\bf n},\varepsilon,{\bf r},t)
+\xi_{\text {c}}^{{\text {sf}}}({\bf n},\varepsilon,{\bf r},t)].
\label{df0}
\end{eqnarray}
In the same way from Eqs. (\ref{fnex}) 
and (\ref{BLfp}), for $f_{\text {s}}$, we obtain
\begin{eqnarray}
&&
\frac{v_{F}}{3}\nabla . {\bf f}_{\text {s}1}=-\frac{1}{\tau _{0}}
f_{\text {s}0}+\int d{\bf n}\xi^{{\text {sf}}}_{\text {s}}({\bf n}
,\varepsilon,{\bf r},t),
\label{dfp}\\*
&&
\frac{v_{F}}{3}\nabla f_{\text {s}0}=-\frac{{\bf f}_{\text {s}1}}
{3\tau_{s}}\nonumber
\\*
&&+\int {\bf n}d{\bf n}[\xi^{{\text {imp}}}_{\text
{s}}({\bf n},\varepsilon,{\bf r},t) +\xi^{{\text {sf}}}_{\text
{s}}({\bf n},\varepsilon,{\bf r},t)],
\label{dfp0}
\end{eqnarray}
where different relaxation times of normal impurity  and spin-flip 
scatterings are defined as
\begin{eqnarray}
&&
\frac{1}{\tau_{{\text {c}}(\text s)}}=\frac{1}{\tau _{{\text
{imp}}}}+\frac{1}{\tau^{-(+)}_{{\text {sf}}}},
\\*
&&
\frac{1}{\tau_{\text {sf}}}=\frac{1}{\tau^{-}_{{\text
{sf}}}}+\frac{1}{\tau^{+}_{{\text {sf}}}},
\label{taus}
\\*
&&
\frac{{\bf n}}{\tau _{{\text {imp}}}}= \int d{\bf
n}^{\prime} w^{{\text {imp}}}({\bf n}, {\bf
n}^{\prime},{\bf r}) ({\bf n}-{\bf n}^{\prime}),
\label{tauimsf}\\*
&&
\frac{{\bf n}}{\tau ^{\pm} _{{\text {sf}}}}= \int d{\bf
n}^{\prime} w^{\text {sf}}({\bf n},{\bf n}^{\prime},{\bf r}) ({\bf
n}\pm {\bf n}^{\prime}).
\label{tausfp}
\end{eqnarray}
Here we used the identity 
$
w^{{\text {imp}}({\text {sf}})}({\bf n},{\bf n}^{\prime}
,{\bf r})= w^{{\text{imp}}({\text {sf}})}(|{\bf n}-{\bf n}
^{\prime}|,{\bf r}), 
$ for the elastic scattering.
\par
In obtaining Eqs.~(\ref{df}-\ref{dfp0}) we have disregarded terms
$(\partial / \partial t)f_{{\text {c}}(\text {s})}$ in the expressions
$(d/dt)f_{{\text {c}}(\text {s})}
=(\partial /\partial t +v_{F}{\bf n}.\nabla +e{\bf
E}.\nabla_{{\bf p}})f_{{\text {c}}(\text {s})}$, since we are only 
interested in the zero frequency noise power. The terms of the electric 
field $e{\bf E}.\nabla_{{\bf p}}$ are eliminated by substituting 
$\varepsilon$ by $\varepsilon-e\varphi_{{\text {c}}({\text {s}})}
({\bf r},t)$ in the arguments of $f_{{\text {c}}({\text {s}})}$, 
respectively, where the charge and spin potentials are expressed 
as $\varphi_{\text {c}}=\sum _{\alpha} \varphi_{\alpha}/2$ and 
$\varphi_{\text {s}}=\sum _{\alpha}\alpha \varphi_{\alpha}/2$, with 
$ \varphi_{\alpha}({\bf r},t)=
\int d\varepsilon {f}_{\alpha 0} (\varepsilon,{\bf r},t),$ being the
spin-dependent electro-chemical potential. We also used the
identities $\int d{\bf n}\xi^{{\text {imp}}}_{{\text {c}}({\text {s}})}
=0,$ which follows from the conservation of number of spin $\alpha$ 
electrons in each normal impurity scattering process\cite{blanterRev}.  
In contrast to this, we note that the integral $\int d{\bf n}\xi^{{\text
    {sf}}}_{{\text {c}}(\text s)}$ does not vanish, reflecting the fact 
that spin is not conserved by the spin-flip process.
\par
Combining  Eqs. (\ref{df}), (\ref{df0}) and (\ref{dfp}),
 (\ref{dfp0}) the equations for the symmetric parts of
the mean charge and spin distribution functions are obtained as
\begin{eqnarray}
&&
{\nabla}^2 {\bar f}_{{\text {c}}0}=0,
\label{delf}\\*
&&
{\nabla}^2 {\bar f}_{\text {s}0}= \frac{{\bar f}_{{\text
s}0}}{{\ell^{2}_{\text {sf}}}},
\label{delfs}
\end{eqnarray}
where 
$\ell_{\text {sf}}=\sqrt{D_{\text {s}}\tau_{\text {sf}}}$ 
is the spin-flip length. Note, that in general charge and spin 
diffusion constants given by  
$
D_{{\text {c}}({\text {s}})}=
v_{F}^2\tau_{{\text {c}}({\text {s}})}/3
$
are different.
\par
Using Eqs. (\ref{fnex}) the charge and spin current
densities can be expressed as 
$
{\bf j}_{{\text {c}}(\text{s})}=(eN_{0} v_{\text F}/3)\int d\varepsilon 
{\bf f}_{{\text {c}}({\text {s}})1}
$. The corresponding fluctuating potentials are given by 
$
{\bar \varphi}_{{\text {c}}(\text {s})}({\bf r})+\delta 
\varphi_{{\text {c}}(\text {s})}({\bf r},t) 
=(1/e)\int d\varepsilon f_{{\text {c}}({\text {s}})0}$. 
Using these identities and integrating Eqs. (\ref{df}-\ref{df0})
over $\varepsilon$ we obtain diffusion equations for the charge
potential and current density,
\begin{eqnarray}
&&
\nabla . {\bar {\bf j}}_{\text {c}}=0,\\*
&&
\nabla . \delta{\bf j}_{\text {c}}=i_{\text {c}}^{{\text {sf}}} 
\label{dj}
\\*
&&
{\bar {\bf j}}_{\text {c}}=-\sigma \nabla {\bar \varphi}_{\text {c}},
\label{barj}
\\*
&&
\delta {\bf j}_{\text {c}}=-\sigma\nabla \delta \varphi_{\text {c}} 
+{\bf j}_{\text {c}}^{c}, 
\label{j}
\end{eqnarray}
which also imply
\begin{equation}
{\nabla}^2 {\bar \varphi}_{\text {c}}=0.
\label{del2fi}
\end{equation}
In the same way Eqs. (\ref{dfp}-\ref{dfp0}) give us diffusion 
equations of spin potential and current density:
\begin{eqnarray}
&&
\nabla . {\bar {\bf j}}_{\text {s}}=-\frac{e^2 \nu _{F}}{\tau
_{{\text {sf}}}} {\bar \varphi }_{\text {s}}, \\*
&&\nabla . \delta {\bf j}_{\text {s}}=-\frac{e^2 \nu _{F}}
{\tau _{{\text {sf}}}} \delta \varphi
_{\text {s}}+i^{{\text {sf}}}_{{\text {s}}}, \label{djp}
\\*
&&
{\bar {\bf j}}_{\text {s}}=-\sigma_{{\text {s}}} \nabla {\bar
\varphi} _{\text {s}},
\label{barjp}
\\*
&&
\delta{\bf  j}_{\text {s}}=
-\sigma_{{\text {s}}}\nabla \delta \varphi_{{\text {s}}} +{\bf
j}_{\text {s}}^{c}, \label{deljp}
\\*
&&
{\nabla}^2 {\bar \varphi}_{{\text {s}}}=
\frac{{\bar \varphi}_{{\text {s}}}}{{\ell^{2}_{\text {sf}}}},
\label{del2fip}
\end{eqnarray}
where $ \sigma_{{\text {c}}({\text {s}})}=
e^2 N_{0} D_{{\text {c}}({\text {s}})}$ are the
charge and spin conductivities. Here
\begin{eqnarray}
{\bf j}^{c}_{{\text {c}}({\text {s}})}=
ev_{F}N_{0}\tau_{{\text {c}}({\text {s}})} \int
(\xi^{{\text {imp}}}_{{\text {c}}({\text {s}})}
+\xi_{{\text {c}}({\text {s}})}^{\text
{sf}}){\bf n}d{\bf n}d\varepsilon \label{jcs}
\end{eqnarray}
are the Langevin sources of fluctuations of the charge and spin 
current densities, and
\begin{eqnarray}
i^{{\text {sf}}}_{{\text {c}}({\text {s}})}({\bf r},t)
=eN_{0}\int d\varepsilon d{\bf n}\xi^{{\text {sf}}}_{{\text {c}}
({\text {s}})}, \label{isp}
\end{eqnarray}
are additional terms in the expressions for the divergence of charge and
spin currents fluctuations Eqs.~(\ref{dj}) and (\ref{djp}), due to the
non-conserved nature of spin-flip process.
\par
Now we calculate possible correlations between the currents 
${\bf j}_{{\text {c}}(\text {s})}^{c}$  
and 
$i^{{\text {sf}}}_{{\text {c}}{(\text {s})}}$.
From Eqs. (\ref{ksis}-\ref{poisson}) for the correlations
of different fluctuating sources we obtain
\begin{eqnarray}
&&<\xi^{{\text {imp}}({\text {sf}})}_{\alpha}({\bf
n},\varepsilon,{\bf r},t) \xi^{{\text {imp}}({\text
{sf}})}_{\alpha^{\prime}}({\bf n}^{\prime} ,\varepsilon^{\prime},{\bf
r}^{\prime},t^{\prime})>= \frac{1}{N_{0}}\nonumber \\* &&
\times \delta({\bf r}-{\bf r}^{\prime})
\delta(t-t^{\prime})\delta(\varepsilon-\varepsilon^{\prime}) G^{{\text
{imp}}({\text {sf}})}_{\alpha \alpha^{\prime}} ({\bf n},{\bf
n}^{\prime},{\bf r},\varepsilon),\label{ksicor}\\*
 &&<\xi^{\text
{imp}}_{\alpha}({\bf n},\varepsilon,{\bf r},t) \xi^{\text
{sf}}_{\alpha^{\prime}}({\bf n}^{\prime} ,\varepsilon^{\prime},{\bf
r}^{\prime},t^{\prime})>=0,\label{ksicor0}
\end{eqnarray}
where
\begin{eqnarray}
&& G^{{\text {imp}}}_{\alpha\alpha^{\prime}}= \delta
_{\alpha\alpha^{\prime}}\int d{\bf n} ^{\prime \prime}[\delta({\bf
n}-{\bf n}^{\prime})- \delta({\bf n}^{\prime}-{\bf n}^{\prime
\prime})] \nonumber \\* && [{\bar J}_{\alpha \alpha}({\bf n},{\bf
n}^{\prime \prime},\varepsilon)+ {\bar J}_{\alpha \alpha}({\bf
n}^{\prime \prime},{\bf n},\varepsilon)], \label{Gssp}
\\* 
&&
G^{\text {sf}}_{\alpha\alpha^{\prime}}= 
\int d{\bf n}^{\prime \prime}[\delta _{\alpha\alpha^{\prime}}
\delta({\bf n}-{\bf n}^{\prime})-
\delta _{\alpha-\alpha^{\prime}} \delta({\bf n}^{\prime}-{\bf
n}^{\prime \prime})]\nonumber \\* && [{\bar J}_{-\alpha
\alpha}({\bf n},{\bf n}^{\prime \prime})+ {\bar J}_{\alpha
-\alpha}({\bf n}^{\prime \prime},{\bf n})]. \label{Gsfssp}
\end{eqnarray}
\begin{figure}
\centerline{\hbox{\epsfxsize=3.in \epsffile{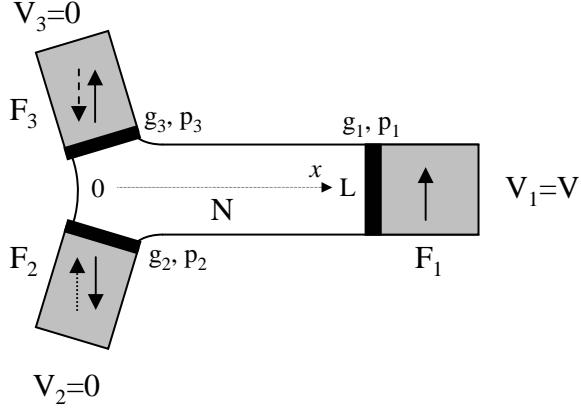}}}
\caption{A schematic picture of the three terminal spin valve. A
diffusive wire of length $L$ is connected to three ferromagnetic
terminals via the tunnel junctions.} \label{zbfig1}
\end{figure}

From Eqs. (\ref{ksiimsf}), (\ref{ksiimsfs}) and 
(\ref{ksicor}-\ref{Gsfssp}) we calculate the correlations 
between the fluctuating sources 
$\xi^{{\text{imp}}({\text {sf}})}_{{\text {c}}(\text {s})}$, 
which can be used together with Eqs. (\ref{jcs}) and 
(\ref{isp}) to obtain the results,
\begin{eqnarray}
&&<j_{{\text {c}}({\text {s}})l}^{c}({\bf r},t) 
j_{{\text {c}}({\text
{s}})m}^{c}({\bf r}^{\prime},t^{\prime})>= 
\delta_{lm} \delta({\bf r}-{\bf r}^{\prime})
\delta(t-t^{\prime}) \sigma_{{\text {c}}({\text {s}})}
\nonumber\\*
&& \times \frac{\tau_{{\text {c}}({\text {s}})}}{\tau_{\text {imp}}} 
\sum_{\alpha}[ \Pi_{\alpha
\alpha}({\bf r}) +\frac{\tau_{\text {imp}}}{\tau^{-(+)}_{\text
{sf}}} \Pi_{\alpha -\alpha}({\bf r})],
\label{jsjs}
\\*
&&<j_{{\text {c}}l}^{c}({\bf r},t) j_{\text {s}m}^{c}({\bf r}^{\prime},
t^{\prime})>=
\delta_{lm} \delta({\bf r}-{\bf
r}^{\prime}) \delta(t-t^{\prime})\sigma_{\text {c}} 
\nonumber 
\\* 
&&
\times \sum _{\alpha}\alpha\Pi_{\alpha \alpha}({\bf r}),
\label{jsjsp}
\\*
&&<i_{\text {c}}^{\text {sf}}({\bf
r},t) i_{{\text {c}}({\text {s}})}^{\text {sf}}({\bf
r}^{\prime},t^{\prime})>\nonumber
\\*  
&&=<j_{{\text {c}}(\text {s})m}^{c}({\bf r}^{\prime},
t^{\prime}) i_{{\text {c}}(\text {s})}^{\text {sf}}({\bf
r}^{\prime},t^{\prime})> =0,
\label{iis}
\\*
&&<i_{\text {s}}^{\text {sf}}({\bf r},t) i_{\text {s}}^{\text
{sf}}({\bf r}^{\prime},t^{\prime})> =\delta_{lm} 
\delta({\bf r}-{\bf r}^{\prime})\delta(t-t^{\prime})\sigma_{\text {c}}
\\* 
&&
\times \frac{1}{D_{\text {c}}\tau^{+}_{\text {sf}}}
\sum _{\alpha} \Pi_{\alpha -\alpha}({\bf r}),
\label{isis}
\end{eqnarray}
where

\begin{equation}
\Pi_{\alpha \alpha^{\prime}}({\bf r})= \int d\varepsilon {\bar
f}_{\alpha 0}(\varepsilon,{\bf r}) [1-{\bar f}_{\alpha^{\prime}
0}(\varepsilon,{\bf r})]. \label{pi}
\end{equation}
The diffusion equations (\ref{delf}-\ref{del2fip}) and Eqs.
(\ref{jsjs}-\ref{pi}) are a complete set of equations, which in
principal can be solved for a multi-terminal mesoscopic diffusive
conductor connected to an arbitrary number $N$ of metallic and/or
ferromagnetic terminals held at constant potentials. The
distribution function $f_{n}(\varepsilon-eV_n)$ of the terminal $n
(=1,...,N)$ biased at the voltage $V_n$ determines the boundary
conditions of the diffusive equations. In the case where the terminal
$n$ is connected by a tunnel junction to the diffusive conductor at the
point ${\bf r}_n$, the ferromagnetic character of the terminal can be
modeled by a spin-dependent conductance $g_{n\alpha}$. The fluctuating
spin $\alpha$ current through the junction is given by
$I_{n\alpha}(t)=g_{n\alpha}\int d\varepsilon [{f}_{\alpha
  0}(\varepsilon,{\bf r}_n,t) -f_{n}(\varepsilon-eV_n)].  $ As the
boundary condition this current should match to the value calculated
from the diffusive equations $I_{n\alpha}=\int_{A_{n}}d{\bf S}\cdot{\bf
  j}_{\alpha}({\bf r}_n,t)$, where $A_n$ is the junction area. Here
${f}_{\alpha 0}={f}_{{\text {c}}0}+\alpha {f}_{{\text {s}}0}$ and ${\bf
  j}_{\alpha}={\bf j}_{\text {c}}+\alpha {\bf j}_{\text {s}}$ are the 
spin $\alpha$ symmetric part of the distribution function and current 
density respectively. From the solutions of the diffusion equations the mean
charge and spin currents and the correlations of the corresponding
fluctuations can be obtained.
\par
Eqs. (\ref{delf}-\ref{del2fip}) and (\ref{jsjs}-\ref{pi}) 
are valid for an arbitrary $\tau_{\text {imp}}/\tau_{\text
{sf}}$ in the diffusive limit. In the following we will 
consider the more realistic case of $\tau_{\text {imp}}\ll \tau_{\text
{sf}}$, where the effect of spin-flip scattering on the conductivity
of the diffusive metal is neglected. In this case,  
$\tau_{\text {s}}=\tau_{\text {c}}=\tau_{\text {imp}}$  and hence 
$\sigma_{\text {s}}=\sigma_{\text {c}}=\sigma$.  
For simplicity, we also assume that
the spin-flip scattering is isotropic, i. e., $w^{{\text
{sf}}}$ does not depend on the directions ${\bf n}$, ${\bf
n}^{\prime}$, which implies  
$\tau^{-}_{\text {sf}}=\tau^{+}_{\text {sf}}
=2\tau_{\text {sf}}$. In the next section we use the above 
developed formalism to calculate spin-polarized current 
correlations in a diffusive three-terminal system.    
\par

\section{Three-terminal spin valve}

We consider the three-terminal spin valve system as shown in Fig.
\ref{zbfig1}. The system consists of a diffusive normal wire (N)
of length $L$ connected by tunnel junctions to three ferromagnetic 
terminals F$_{i}$ ($i=1,2,3$). The terminal F$_{1}$ is held at
the voltage $V$ and the voltage at the terminals F$_{2,3}$ is
zero. The tunnel junction $i$ connecting F$_{i}$ to N has a spin
dependent tunneling conductances $g_{i\alpha}$, which
equivalently can be characterized by a total conductance
$g_{i}=\sum _{\alpha} g_{i\alpha}$, and the polarization
$p_{i}=\sum _{\alpha} \alpha g_{i\alpha}/g_{i}$. Inside the wire we
account for both, normal
impurity and spin flip scattering. The
length $L$ is much larger than $\ell_{{\text {imp}}}$ providing a
diffusive motion of electrons. The spin-flip length $\ell_{{\text
{sf}}}$ is assumed to be much larger than  $\ell_{\text
{imp}}$, but arbitrary as compared to $L$. We study the influence 
of the spin-flip scattering on shot noise of the current through 
the wire and cross-correlations between currents through the 
terminals F$_{2}$ and F$_{3}$.

\subsection{Charge and spin currents fluctuations}
To start we write the solutions of Eqs. (\ref{del2fi}) and
(\ref{del2fip}) in terms of the charge (spin) potentials 
$\varphi_{{\text {c}}({\text s})}(0)$ and 
$\varphi_{{\text {c}}({\text s})}(L)$, 
at the connecting points $x=0$ and $x=L$ inside the  wire:
\begin{eqnarray}
{\bar \varphi}_{{\text {c}}({\text s})} (x) = 
\phi_{{\text {c}}({\text s})0} (x){\bar
\varphi}_{{\text {c}}({\text s})}(0) 
+\phi_{{\text {c}}({\text s})L} (x){\bar
\varphi}_{{\text {c}}({\text s})}(L), 
\label{fi}
\end{eqnarray}
where the charge and spin potential functions are defined as
\begin{eqnarray}
&&
\phi_{{\text {c}}0} (x)=1-\frac{x}{L}, \\*
&&
\phi_{{\text {c}}L} (x)=\frac{x}{L}, \label{pf}
\\*
&&
\phi_{{\text s}0}
(x)=\frac{\sinh{[\lambda(1-x/L)]}}{\sinh{\lambda}},
\label{pfs0}
\\*
&&
\phi_{{\text s}L} (x)=\frac{\sinh{(\lambda x/L)}}{\sinh{\lambda}}.
\label{pfsL}
\end{eqnarray}
Here the parameter $\lambda=L/\ell_{\text {sf}}$ is the 
dimensionless measure of the spin-flip scattering inside 
the N-wire. An expression for the fluctuations of the current 
through the wire $\Delta I_{{\text {c}}1}$ is obtained if we take 
the inner product of $\delta {\bf j}_{\text {c}}$ in Eq. (\ref{j}) 
with $\nabla \phi_{{\text {c}}0}$ 
and integrate over the volume of the wire: 
\begin{eqnarray}
&&
\Delta I_{{\text {c}}1}=-\sigma\int d{\bf s}\cdot\nabla \phi_{{\text {c}}0}
\delta \varphi_{\text {c}}
+\delta I_{\text {c}}^{c}, \label{deli1}
\\*
&&
\delta I_{\text {c}}^c=\int d\Omega (i_{\text {c}}^{\text {sf}}
+{\bf j}_{\text {c}}^c\cdot\nabla)\phi_{{\text {c}}0},
\label{delic}
\end{eqnarray}
where we also used Eqs. (\ref{dj}) and (\ref{del2fi}).  In a similar way
by volume integration of the products $\nabla \phi_{{\text s}0(L)}
\cdot \delta {\bf j}_{\text s}$ in Eq. (\ref{deljp}), and using Eqs.
(\ref{djp}) and (\ref{del2fip}) we obtain $\Delta I_{{\text s}1}(0,L)$,
which yields the fluctuations of spin currents at $x=0$ and $x=L$:
\begin{eqnarray}
&&
\Delta I_{{\text s}1}(0,L)=-\sigma\int d{\bf s}\cdot\nabla
\phi_{{\text {s}}(0,L)} \delta \varphi_{\text {s}} 
 \nonumber \\*
&& +\delta
I^{c}_{\text s}(0,L),
\label{delis1}
\\*
&&
\delta I^c_{\text s}(0,L)= \int d\Omega 
(i^{\text {sf}}_{\text s}+{\bf
j}^c_{\text s}\cdot\nabla)\phi_{{\text s}0,L}. 
\label{delics}
\end{eqnarray}
Note, that as a result of the spin-flip scattering the spin current and,
hence, its fluctuations are not conserved through the wire. Using Eqs.
(\ref{delis1}), (\ref{delics}), (\ref{pfs0}) and (\ref{pfsL}) a relation
between the fluctuation of the spin currents at the two different points
is obtained,
\begin{eqnarray}
&&
\Delta I_{{\text s}1}(0)=\Delta I_{{\text
s}1}(L)-g_{N}\lambda^2t [\delta \varphi_{{\text s}}(0) 
+\delta \varphi_{{\text s}}(L)]
\nonumber\\* 
&& 
+\delta I^{c}_{\text s}(0) 
+\delta I^{c}_{\text s}(L),
\label{delis0l}
\end{eqnarray}
where $g_N=\sigma A/L$ ($A$ being the area of the wire) is the
conductance of the wire and $t(\lambda)=\tanh{\lambda}/\lambda$.
In the limit $\lambda \rightarrow 0$, the conservation of the spin 
current is retained and $\Delta I_{{\text s}1}(0)
=\Delta I_{{\text s}1}(L)$, as is seen in Eq. (\ref{delis0l}).
\par
In the Boltzmann-Langevin formalism, the fluctuation of spin $\alpha$ 
currents through junction $i$ are written in terms of the intrinsic 
current fluctuations $\delta I_{i\alpha}$ due to the random scattering 
of electrons from the tunnel barriers and the potential fluctuations
$\delta \varphi_{\alpha}(0,L)$ at the junction points:
\begin{eqnarray}
&&\Delta I_{i\alpha}(0,L)=\delta I_{i\alpha}-g_{i}\delta 
\varphi_{\alpha}(0,L).
\label{deliia}
\end{eqnarray}
Using this relation the fluctuations of charge and spin currents 
through the terminals can be expressed in terms of the fluctuating 
spin and charge  potentials at the connection points and 
the corresponding intrinsic currents fluctuations. Denoting 
$\delta I_{{\text {c}}({\text s})i}$ as the intrinsic fluctuations of the 
charge (spin) current through the tunnel junction $i$, we obtain
\begin{eqnarray}
&&\Delta I_{{\text {c}}1}=\delta I_{{\text {c}}1}-g_{1}
\delta \varphi_{\text {c}}(L)
-g_{1}p_{1}\delta \varphi_{\text {s}}(L), 
\label{deli1n}\\*
&&\Delta I_{{\text {c}}2,3}=\delta I_{{\text {c}}2,3}-g_{2,3}
\delta \varphi_{\text {c}}(0)
-g_{2,3}p_{2,3}\delta \varphi_{{\text s}}(0), \label{delii}\\*
&&\Delta I_{{\text s}1}(L)=\delta I_{{\text s}1} -g_{1}p_{1}\delta
\varphi_{\text {c}}(L) -g_{1}\delta \varphi_{{\text s}}(L),
\label{delis1n}\\*
&&\Delta I_{{\text s}2,3}=\delta I_{{\text
s}2,3} -g_{2,3}p_{2,3}\delta \varphi_{\text {c}}(0) -g_{2,3}\delta
\varphi_{\text {s}}(0). \label{delisi}
\end{eqnarray}
\par
Now we have to apply the currents conservation rules at the junction 
points. For spin-conserving tunnel barriers charge and spin 
currents fluctuations are conserved. At the pint $x=L$ the rules apply 
as equality of the expressions for $\Delta I_{{\text {c}}1}$ given in Eqs. 
(\ref{deli1}) and (\ref{deli1n}), and $\Delta I_{{\text s}1}(L)$ in Eqs. 
(\ref{delis1}) and (\ref{delis1n}). At the point $x=0$, they read  
\begin{eqnarray}
&&\sum_{i=1}^{3}\Delta I_{{\text {c}}i}=0, 
\label{sumdeli}
\\*
&&\sum_{i=1}^{3}\Delta I_{{\text s}i}(0)=0, 
\label{sumdelis}
\end{eqnarray}
which in combination with Eqs. (\ref{deli1}-\ref{delis0l}) and
(\ref{deli1n}-\ref{delisi}) lead to  
\begin{eqnarray}
&&\sum_{i}\delta I_{{\text {c}}i}=g_{23}\delta \varphi_{\text {c}}(0)
+g_{1}\delta\varphi_{\text {c}}(L) 
\nonumber \\*
&&
+g_{23}p_{23}\delta 
\varphi_{{\text s}}(0) +g_{1}p_{1}
\delta \varphi_{{\text s}}(L)
\label{sys1}\\*
&&
\sum_{i}\delta I_{{\text s}i}= g_{23}p_{23}\delta
\varphi_{\text {c}}(0)+g_{1}p_{1}\delta \varphi_{\text {c}}(L) 
\nonumber \\*
&&
+(g_{23}+g_Nt)
\delta \varphi_{{\text s}}(0)
+(g_{1}+g_{N}t)\delta\varphi_{{\text s}}(L) 
\nonumber \\*
&&
-\delta I^c_{\text s}(0)
-\delta I^c_{\text s}(L)
\label{sys2} \\*
&&
\delta I_{{\text s}1}=g_{1}p_{1}\delta \varphi_{\text {c}}(L) 
-\frac{g_{N}}{s}\delta \varphi_{{\text s}}(0)
\nonumber \\*
&&
+(g_{1}+g_{N}\frac{\lambda^2}{t})
\delta \varphi_{{\text s}}(L)
-\delta I^c_{\text s}(L) 
\label{sys3} \\*
&&
\delta I_{{\text {c}}1}=-g_{N}\delta \varphi_{\text {c}}(0) 
\nonumber \\*
&&
+(g_{1}+g_{N})\delta \varphi_{\text {c}}(L)
+g_{1}p_{1}\delta \varphi_{{\text s}}(L) 
-\delta I_{\text {c}}^c
\label{sys4}
\end{eqnarray}
The solution of this system of equations gives us the fluctuations
of the potentials in the connecting nodes which can be replaced into 
Eqs. (\ref{deli1n}-\ref{delisi}) to obtain the fluctuations of the
charge and spin currents through different terminals in terms of
$\delta I_{{\text {c}}i}$, $\delta I_{{\text s}i}$, 
$\delta I_{\text {c}}^c$, and $\delta I^c_{\text s}(0,L)$. 
In particular, the fluctuations of the charge currents have the form
\begin{eqnarray}
&&\Delta I_{{\text {c}}i}=\sum_{j=1}^{3}(c_{ij}\delta I_{{\text {c}}j}
+c^{s}_{ij}\delta I_{sj})
\nonumber \\*
&&
+c_{i}\delta I_{\text {c}}^c+
c_{i0}\delta I^c_{\text s}(0)+c_{iL}\delta I^c_{\text s}(L),
\label{delif}
\end{eqnarray}
where $c_{ij}$, $c^{s}_{ij}$, $c_{i}$, $c_{i0}$, and $c_{iL}$
are functions of $g_{i}$, $p_{i}$, and $\lambda$.

\subsection{Mean currents and correlations of currents fluctuations}
The currents correlations  $\langle\Delta I_{{\text {c}}i} 
\Delta I_{{\text {c}}j}\rangle$,
are expressed in terms of the correlations of different fluctuating 
currents appearing in  Eqs. (\ref{delif}).
To calculate the correlations of the currents $\delta I_{\text {c}}^c$ 
and
$\delta I^c_{\text s}(0,L)$ we have to determine the mean
distribution function ${\bar f}_{\alpha 0}={\bar f}_{{\text {c}}0}
+\alpha {\bar
 f}_{{\text s}0}$. This is achieved by solving Eqs. (\ref{delf}),
(\ref{delfs}) and imposing the boundary conditions that ${\bar
f}_{\alpha 0}$ in the terminal F$_i$ held in equilibrium at 
the voltage $V_i$ is given by the Fermi-Dirac distribution function 
$f_i=F_{\text {FD}}(\varepsilon-eV_i)$. From the solutions of 
Eqs. (\ref{delf}), (\ref{delfs}) we obtain
\begin{eqnarray}
&&
{\bar f}_{\alpha}=f_1+(f_2-f_1)[a+b \frac{x}{L}
\nonumber
\\*
&&
+\alpha(c \sinh{\frac{\lambda x}{L}}+
d\cosh{\frac{\lambda x}{L}})], 
\label{fms}
\end{eqnarray}
where $a, b, c, d$ are coefficients which have to be determined by
the boundary conditions. Integrating of (\ref{fms}) over the energy
$\varepsilon$ the mean electro-chemical potential of spin 
$\alpha$ electrons is obtained:
\begin{eqnarray}
&&\nonumber
{\bar \varphi}_{\alpha}(x)=[a+b \frac{x}{L}
\\*
&&
+\alpha(c \sinh{\frac{\lambda x}{L}}+d\cosh{\frac{\lambda x}{L}})]V,
 \label{fims}
\end{eqnarray}
which also could be obtained from the solutions of Eqs.
(\ref{del2fi}), (\ref{del2fip}). 
\par
In the presence of the tunnel junctions the boundary conditions are 
imposed by applying the mean currents conservation rules at the 
connection points. Using Eqs. (\ref{barj}), (\ref{barjp}), (\ref{fims}) 
we obtain the mean charge and spin currents through the N wire:
\begin{eqnarray}
&&
{\bar I}_{{\text {c}}1}=b g_{N}V, 
\label{imc}
\\*
&&
{\bar I}_{{\text s}1}(x)=g_{N}\lambda
(c \sinh{\frac{\lambda x}{L}}+d\cosh{\frac{\lambda x}{L}})V.
\end{eqnarray}
In terms of the charge and spin potentials at the connection points 
${\bar \varphi}_{\text {c}}(0)=a V$, 
${\bar \varphi}_{\text {c}}(L)= (a+bL)V$, 
${\bar \varphi_{{\text s}}}(0)= dV$,
${\bar \varphi_{{\text s}}}(L)= (\lambda sc+\cosh{\lambda}d)V$,
we have the following relations for the mean currents
\begin{eqnarray}
&&
{\bar I}_{{\text {c}}1}=-g_{1}{\bar \varphi}_{\text {c}}(L) -g_{1}p_{1}{\bar
\varphi_{{\text s}}}(L),
\label{i1m}
\\* 
&&{\bar I}_{{\text {c}}2,3}=-g_{2,3}{\bar
\varphi}_{\text {c}}(0) -g_{2,3}p_{2,3}{\bar \varphi_{{\text s}}}(0),
\\*
&&{\bar I}_{s1}=-g_{1}p_{1}{\bar \varphi}_{\text {c}}(L) -g_{1}{\bar
\varphi_{{\text s}}}(L),
\\* 
&&{\bar I}_{s2,3}=-g_{2,3}p_{2,3}{\bar
\varphi}_{\text {c}}(0) -g_{2,3}{\bar \varphi_{{\text s}}}(0), 
\label{iim}.
\end{eqnarray}
Using Eqs. (\ref{imc}-\ref{iim}) and the currents conservations
relations at the point $x=0$, $\sum_{i=1}^{3}{\bar I}_{{\text {c}}i}=0$, 
and $\sum_{i=1}^{3}{\bar I}_{{\text s}i}(0)=0$, we find
\begin{eqnarray}
&&\frac{a}{C}=[\frac{g_N}{g_{23}}
+(1+\frac{g_N}{g_{1}})q_{23}
+\frac{g_1}{g_{23}}q_{1}]\cosh{\lambda}
\nonumber
\\*
&&
+[\frac{g_N}{g_{23}}(1+\frac{g_N}{g_{1}})\lambda^2 +
(\frac{g_1}{g_{N}}+q_{1})q_{23}]s\nonumber
\\*
&&
-p_1p_{23}\frac{g_N}{g_{23}},
\label{aa}
\\*
&&
\frac{1}{C}=
[2\frac{g_N}{g_{23}}+(1+\frac{g_N}{g_{1}})q_{23}
+\frac{g_1}{g_{23}} (1+\frac{g_N}{g_{23}})
q_{1}]
\nonumber
\\*
&&
\times \cosh{\lambda} 
+\frac{g_N}{g_{23}}(1+\frac{g_N}{g_{1}}+\frac{g_N}{g_{23}})
\lambda^2s
\nonumber  
\\*
&&
+[q_{1}+q_{23}(1+\frac{g_1}{g_{N}}q_{1})]s
-2p_1p_{23},
\label{ccc}
\\*
&&
\frac{b}{C}=  
-(q_{23}+\frac{g_1}{g_{23}}q_{1})\cosh{\lambda}
\nonumber  
\\* 
&&
-[\frac{g_N}{g_{23}}\lambda^2+\frac{g_1}{g_{N}}
q_{1}q_{23}]s,
\label{bb}
\\*
&& 
\frac{c}{C}
=-\frac{g_1}{g_{23}}(\frac{g_N}{g_{1}}\lambda^2s
+q_1\cosh{\lambda})p_{23}-q_{23}p_1,
\\*
&&
\frac{d}{C}=\frac{g_1}{g_{23}}\lambda(\frac{g_N}{g_{1}}
\cosh{\lambda}+q_1 s)p_{23}
-\frac{g_N}{g_{23}}\lambda p_1,
\label{dd}
\end{eqnarray}
where $p_{23}=(g_2p_2+g_3p_3)/g_{23}$,
$g_{23}=g_2+g_3$,$q_1=1-p_1^2$, $q_{23}=1-p_{23}^2$,
$s(\lambda)=\sinh{\lambda}/\lambda$.
\par
Replacing ${\bar f}_{\alpha 0}$ given by Eq. (\ref{fms}) in Eqs. (\ref{pi})
we can calculate the correlations in Eqs. (\ref{jsjs}-\ref{isis}), which 
can be used to calculate all the possible correlations between 
$\delta I_{\text {c}}^c$ and $\delta I^c_{\text s}(0,L)$, given by 
Eqs. (\ref{delic}) and (\ref{delics}). Calculations lead
to the following results
\begin{eqnarray}
&&
S_{\text {c}}=\langle\delta I_{\text {c}}^{c}\delta I_{\text {c}}^{c}\rangle=
2g_N[a(1-a-b)+\frac{b}{2}(1-\frac{2}{3}b)
\nonumber
\\*
&& 
+\frac{c^2-d^2}{2}-s(cd\lambda s
+\frac{c^2+d^2}{2}\cosh{\lambda})] 
\label{S},\\* 
&&
S_{\text s}(0)= \langle\delta
I^{c}_{\text s}(0)\delta I^{c}_{\text s}(0)\rangle=
2g_N[
\frac{a}{t}(1-a) 
\nonumber
\\*
&& 
+\frac{b}{2}(1-2a)
+\frac{1}{2\lambda^2}(\frac{1}{s^2}-\frac{1}{t})b^2
\nonumber
\\*
&& 
+\frac{c^2-d^2}{2s^2}
-\frac{1}{s}(cd\lambda s+\frac{c^2+d^2}{2}\cosh{\lambda})],
\\*
&&S_{\text s}(L)=\langle\delta I^{c}_{\text s}(L)\delta
I^{c}_{\text s}(L)\rangle=
2g_N[ \frac{a}{t}(1-a)
\nonumber
\\*
&& 
+(\frac{1}{t}-\frac{1}{2})b(1-2a)
+(1-\frac{1}{t}+\frac{1/s^2-1/t}{2\lambda^2})b^2
\nonumber
\\*
&& 
+\frac{c^2-d^2}{2s^2}
-\frac{1}{s}(cd\lambda s+\frac{c^2+d^2}{2}\cosh{\lambda})
],\\* &&S_{\text
s}(0L)=\langle\delta I^{c}_{\text s}(0)\delta I^{c}_{\text
s}(L)\rangle=-2g_N[ \frac{a}{t}(1-a) 
\nonumber\\* 
&&
+\frac{b}{2s}(1-2a)
+(\frac{1/t-1}{2\lambda^2s}-\frac{1}{2s})b^2 
+\frac{c^2-d^2}{2st}
\nonumber\\* 
&&
-\frac{1}{t}(cd\lambda s+\frac{c^2+d^2}{2}\cosh{\lambda})
],
\\* &&S(0)=
\langle\delta I_{\text {c}}^{c}\delta I^{c}_{\text s}(0)\rangle=
g_N[(1-2a-2b)d+(1-2a  
\nonumber\\* 
&&
-b)(\lambda c
+\frac{\cosh{\lambda}}{s}d)
+\frac{1}{\lambda}bc+
(d -\frac{\cosh{\lambda}}{\lambda s} c)b],\\* 
&&
S(L)=\langle\delta I_{\text {c}}^{c}\delta
I^{c}_{\text s}(L)\rangle= -g_N[(1-2a-b)
(\lambda sc 
\nonumber\\* &&
+\frac{1+s\cosh{\lambda}}{s}d)
-(\frac{\cosh{\lambda}}{s}-1)
(sd+\frac{\cosh{\lambda}}{\lambda}c)b]. 
\label{SL}
\end{eqnarray}
In writing these equations, we have for simplicity dropped the time
dependence of the correlators and implicitly assumed, that the
correlators are symmetrized (which leads to a factor of 2). Since we
will be solely interested in the zero-frequency noise correlations, we
also dropped the time integration. 
\begin{figure*}
\centerline{\hbox{\epsfxsize=7.in \epsffile{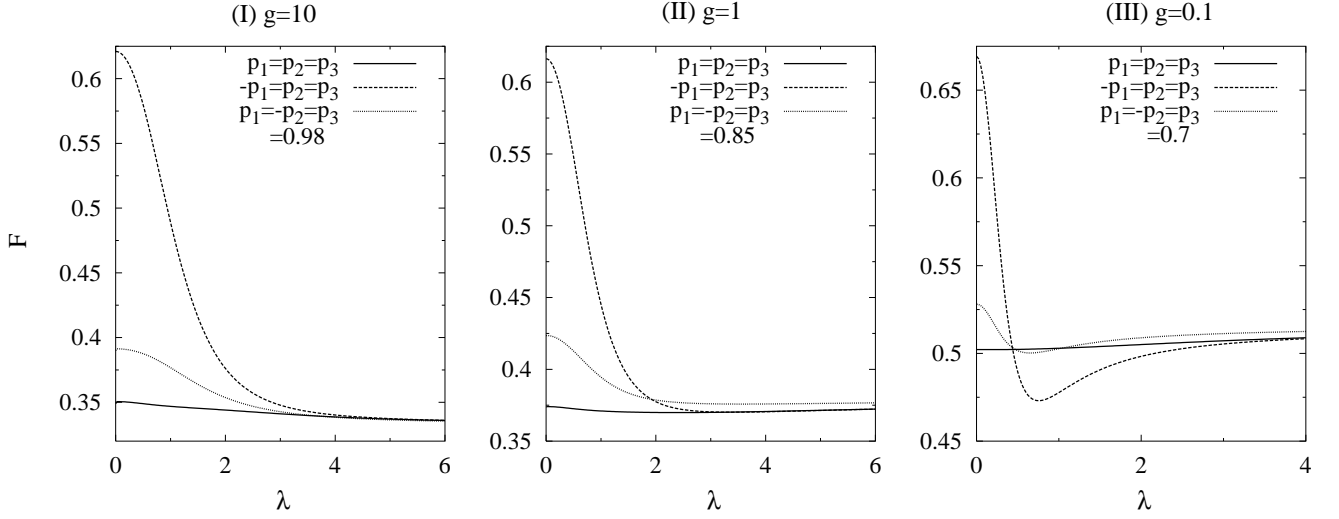} }}
\caption{Fano factor $F$ versus the spin flip scattering intensity
$\lambda=L/\ell_{\text sf}$ for a given magnitude of the polarization 
in F-terminals but different configurations and for different
values of the tunnel conductances $g=g_{i}/g_{N}$. For the 
configurations with $p_2=p_3$ the corresponding cross-correlation is
obtained via $S_{23}/|I_{{\text {c}}1}|=(F-1)/4$.}
\label{zbfig2}
\end{figure*}

To obtain the noise in the terminals we have to specify the correlators
of the intrinsic fluctuations at the tunnel junctions.  Assuming the
tunnel junctions to be spin conserving, we obtain for the correlations
of the intrinsic fluctuation of charge and spin currents
\begin{eqnarray}
&&\langle\delta I_{{\text {c}}i}\delta I_{{\text {c}}j}\rangle=
\langle\delta I_{{\text s}i}\delta
I_{{\text s}j}\rangle =\delta_{ij}2e{\bar
I}_{{\text {c}}i}\label{didi} \\*
&&\langle\delta I_{{\text {c}}i}\delta I_{{\text s}j}\rangle
=\delta_{ij}2e(|{\bar I}_{{\text {c}}i+}|-|{\bar I}_{{\text {c}}i-}|)
.\label{didsj}
\end{eqnarray}
where 
${\bar I}_{{\text {c}}i\alpha}={\bar I}_{{\text {c}}i}
+\alpha {\bar I}_{{\text s}i}$
is the mean current of spin $\alpha$ electrons. 
\par
Using the results (\ref{S}-\ref{didsj}) and Eqs.
(\ref{deli1n}-\ref{delisi}) with the solutions of Eqs.
(\ref{sys1}-\ref{sys4}) the correlation of the currents of the
form 
$S_{ij}=\langle\Delta I_{{\text {c}}i} \Delta I_{{\text {c}}j}\rangle$ 
is obtained. In terms of the coefficients introduced in (\ref{delif}) 
it has the form 
\begin{eqnarray}
&&
S_{ij}=2e\sum_{k=1}^{3}[(c_{ik}c_{jk}+c^{\text s}_{ik}c^{\text s}_{jk})
{\bar I_{k}}+(c_{ik}c^{\text s}_{jk}
\nonumber \\*
&&
+c^{\text s}_{ik}c_{jk})
(|{\bar I_{{\text {c}}+k}}|-|{\bar I_{{\text {c}}-k}}|)]
+c_{i}c_{j}S_{\text {c}}+c_{i0}c_{j0}S_{\text s}(0)
\nonumber \\*
&&
+c_{iL}c_{jL}
S_{\text s}(L)
+(c_{i}c_{j0}+c_{i0}c_{j})S(0)
+(c_{i}c_{jL}
\nonumber \\*
&&
+c_{iL}c_{j})S(L)
+(c_{i0}c_{jL}+c_{iL}c_{j0})S_{\text s}(0L).
\label{sijf}
\end{eqnarray}
In this way we obtain the Fano factor $F=S_{11}/2e|I_{{\text {c}}1}|$ 
and the cross correlations $S_{23}$ measured between the currents 
through F$_2$, F$_3$. In the general case for arbitrary $g_i$, 
$p_{i}$ and $\lambda$
the expressions of $F$ and $S_{23}$ are too lengthy to be given here and
in the next section we will present analytical expressions of $F$
and $S_{23}$ in some important limits only.

\section{Results and discussion}

For simplicity in the following we will consider the 
junctions to have the same tunneling conductances 
$g_{i}/g_{N}=g$, $(i=1,2,3)$. If the amplitude of the 
polarizations $\{p_i\}$ are also the same we distinguish 
the following different configurations. 
The two terminals F$_2$ and F$_3$ have either parallel or 
anti-parallel polarizations. We take the signs of $p_2$ 
and $p_3$ to be positive in the parallel configuration. 
In this case there are two different configurations  
depending on whether $p_1$ is positive (parallel to $p_2$, 
$p_3$) or negative (anti-parallel to $p_2$, $p_3$). 
On the other hand for the antiparallel configuration of F$_2$ 
and F$_3$ the sign of $p_1$ is not essential and the two 
cases of $p_1$ and $-p_1$ are equivalent. Thus there are 
three independent configurations of the polarizations 
corresponding to $p_{1}=p_{2}=p_{3}=p$, $-p_{1}=p_{2}=p_{3}=p$ 
and $p_{1}=-p_{2}=p_{3}=p$. We denote these configurations 
by $"+"$, $"-"$, and $"0"$, respectively.
\par
We analyze the dependence of $F$ and $F_{23}=S_{23}/|I_{{\text {c}}1}|$
on the spin flip scattering intensity $\lambda$ for different 
configurations and amplitude of the polarizations. We show how 
the spin flip scattering affects both $F$ and $S_{23}$ and  
produces a strong dependence on the magnetic configuration 
of the F terminals.

\subsection{Shot noise}

Let us start with analyzing of the shot noise.  Fig. \ref{zbfig2}
illustrates the typical behaviour of the shot noise with respect to the
spin flip scattering intensity and configuration of the polarizations.
Here $F$ versus $\lambda$ is plotted for a given magnitude of the
polarization $p$ in the terminals for the different magnetization
configurations. Different columns I-III belong to different values of
the tunnel contact conductances $g$.  Clearly, for a finite $\lambda$
the Fano factor $F$ is changes drastically with the relative orientation
of the polarizations.  For strong spin flip scattering, $\lambda\gg 1$
the Fano factor reduces to
\begin{eqnarray}
F_{N}=\frac{1}{3}\frac{5+6g+4g^2+(8/9)g^3}{3+6g+4g^2+(8/9)g^3},
\label{fn}
\end{eqnarray} 
independent of the polarization of the terminals. Eq. (\ref{fn}) is the
result for a all-normal metal three terminal system ($p_1=p_2=p_3=0$)
\cite{blanterRev} which reduces to $5/9$ and $1/3$
in the limits of small and large $g$, respectively. This is expected
since the strong spin flip intensity destroys the polarization of the
injected electrons.

At finite $\lambda$ the curves belonging to different configurations
differ from each other and the largest difference occurs as $\lambda$
approaches zero.  In this limit, using Eqs.  (\ref{sijf}) and
(\ref{i1m}) we obtain the following results for the different
configurations
\begin{eqnarray}
&&F_{+}=\frac{45G_{+}^2}{g^2x_{+}^3}[q(\frac{4-2q}{9}g^2
+g+\frac{3}{2})g+\frac{3}{4}] 
\nonumber \\*
&&+\frac{16G_{+}^3}{3x_{+}^4g^2}\{q^3(qg+\frac{21}{2})g^5
+\frac{3}{2}q[q(22+7q)g+36+\frac{51}{2}q]g^3
\nonumber \\*
&& 
-\frac{81}{32}q[(6q-32)g^2+11g+6]
+\frac{243}{8}(\frac{4}{3}g^2+3g+1)
\},
\label{frpp}
\end{eqnarray}
\begin{eqnarray}
&&F_{-}=\frac{G_{-}^2}{q^2g^2x_{-}^3}[
10q^3(2-q)g^3+9q^2(7-2q)g^2+6q(\frac{5}{4}q^2
\nonumber \\*
&&
-6q+16)g+3(\frac{13}{4}q^2-8q+16)]
+\frac{8G_{-}^3}{qx_{-}^4g^2}[
\frac{2}{3}q^5g^6
\nonumber \\*
&&
+7q^4g^5
+q^3(3q+26)g^4
+\frac{3}{2}q^2(9q+32)g^3
\nonumber \\*
&&
+\frac{1}{3}q(-\frac{83}{8}q^2
+86q+137)g^2
+9(-\frac{17}{16}q^2+\frac{15}{4}q+2)g
\nonumber \\*
&&
+\frac{q^2}{8}-7q+17],
\label{frpm}
\end{eqnarray}
\begin{eqnarray}
&&F_{0}=\frac{2G_{0}^2}{g^2x_{0}^3}[q(q^2-4q+8)g^3+3q(\frac{q^2}{2}-q+8)g^2
\nonumber \\*
&&
+3q(\frac{q^2}{4}-q+12)g
+\frac{q^3}{8}-\frac{5}{4}q^2+10q
\nonumber \\*
&&
+8]
+\frac{8G_{0}^3}{x_{0}^4g^2}[
\frac{2}{3}q^4g^6+\frac{7}{3}q^3(q+2)g^5+
q^2(3q^2
\nonumber \\*
&&
+18q+8)g^4
+q(\frac{11}{6}q^3+25q^2+24q+\frac{32}{3})g^3+\frac{1}{3}(\frac{13q^4}{8}
\nonumber \\*
&&
+55q^3+44q^2+96q+16)g^2
+(\frac{q^4}{16}+\frac{57}{8}q^3+q^2
\nonumber \\*
&&
+\frac{57q}{4}+16)g
+\frac{9}{8}q^3-q^2+2q+8]\,.
\label{frpz}
\end{eqnarray}
Here we defined  $x_{\pm}=2qg+3$, $x_{0}=2qg+q+2$, and $q=1-p^2$.  
The total conductances of the system 
normalized by $g_N$ for the three configurations
in the limit $\lambda \rightarrow 0$ are given by
\begin{eqnarray}
  \nonumber
  G_{+} & = & \frac{gx_{+}}{2qg^2+6g+9/2}\,,\\\nonumber
  G_{-} & = & \frac{2qgx_{-}}{4q^2g^2+12qg+q+8}\,,\\\nonumber
  G_{0} & = & \frac{gx_{0}}{2qg^2+2qg+4g+q/2+4}\,.
\end{eqnarray}

\begin{figure}
\centerline{\hbox{\epsfxsize=3.20in \epsffile{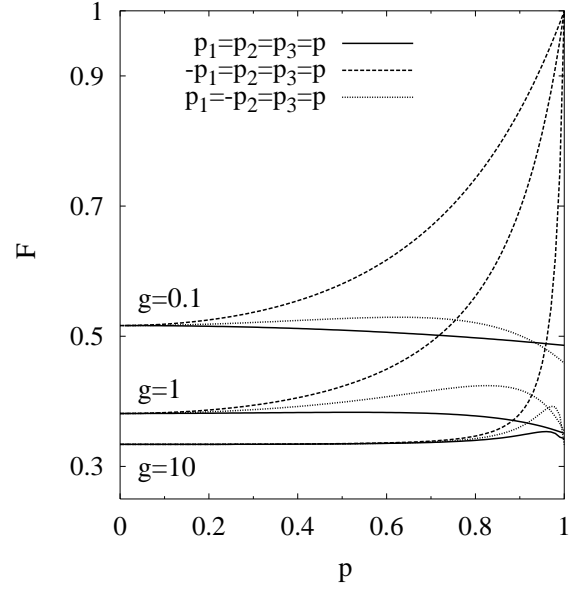} }}
\caption{Fano factor $F$ versus the magnitude of spin polarization
$p$ of the terminals in the limit of small spin flip-scattering 
intensity $\lambda=L/\ell_{\text sf} \rightarrow 0$. The results 
are shown for different configurations of the polarizations and 
different values of the tunnel conductances $g$.}
\label{zbfig3}
\end{figure}
In Fig. \ref{zbfig3} we show the polarization dependence of the Fano
factor in the limit of small spin-flip scattering intensity, $\lambda
\rightarrow 0$ for the different magnetic configurations of the
terminals. At $p=0$ the Fano factor for different conductances $g$ takes
the normal state value $F_N$, see Eq. (\ref{fn}), independent of the
polarizations configuration. For finite polarizations $F$ of different
configurations differ from each other and the normal state value. As $p$
increases, the Fano factor of the configuration $-$ deviates
substantially from those of the other two configurations and reaches the
full Poissonian value $1$ as $p$ approaches $1$. This is independent of
the tunnel conductances $g$.  Thus for perfectly polarized terminals the
Fano factor takes the value $1$ in the limit of small spin-flip
scattering rate.
\par
To understand this effect we note that in the limit $p\to 1$, the system
constitutes an ideal three terminal spin valve due to the antiparallel
configuration of the polarizations at its two ends. In the absence of
the spin flip scattering there in no current through the N-wire since
for the up-spin electrons provided by the terminals F$_2$ and F$_3$,
there is no empty state in the terminal F$_1$ in the energy range $eV$.
For very small but finite $\lambda$ only those of electrons which
undergo spin-flip scattering once can carry a small amount of current.
These spin-flipped electrons are almost uncorrelated and pass through
the normal wire independently giving rise to full Poissonian shot noise.
Similar effects have been found before for two \cite{mishchenko03} and
four \cite{BZ04} terminal spin-valve systems.

\par
For arbitrary $g$ and $p$, $F$  has a complicated dependence 
on $\lambda$ and the corresponding expressions are to lengthy to be
presented here. Simpler expressions are obtained for perfectly polarized
junctions. Setting $p=1$ in Eqs. (\ref{sijf}) and (\ref{i1m}) yield for
the Fano factors of the different configurations 
\begin{eqnarray}
&&F_{+}=\frac{5G_{+}^2}{g^2} \nonumber \\*
&&+\frac{8G_{+}^3}{g^2}[\frac{1}{3}g^2+\frac{3}{2}g+1
+2(g^2+\frac{2g^2}{\lambda^2}+3g+1)
\frac{\tanh{(\lambda/2)}}{\lambda}
\nonumber \\*
&&+4g(g\frac{\cosh^2{\lambda}-3s}
{ \lambda^2s^2}+\frac{3}{2})\frac{\tanh^2
{(\lambda/2)}}{\lambda^2}],
\label{flpp}
\end{eqnarray}
\begin{eqnarray}
&&F_{-}=\frac{5G_{-}^2}{g^2} \nonumber \\*
&&+\frac{8G_{-}^3}{g^2}[\frac{1}{3}g^2+\frac{3}{2}g+1
+2(g^2\frac{+2g^2}{\lambda^2}+3g+1)
\frac{\coth{(\lambda/2)}}{\lambda}
\nonumber \\*
&&+4g(g\frac{\cosh^2{\lambda}+3s}
{\lambda^2s^2}+\frac{3}{2})\frac{\coth^2{(\lambda/2)}}
{\lambda^2}],
\label{flpm}
\end{eqnarray}
\begin{eqnarray}
&&F_{0}=\frac{G_{0}^2}{g^2}[g^2(1+\frac{1}{\lambda^2})
+3g+\frac{9}{4}-\frac{3G_{0}}{g}(\frac{2}{9}g^3+g^2
+\frac{3}{2}g
\nonumber \\*
&&+\frac{1}{2})]
-\frac{G_{0}}{x^2}[4g^2+2g-\lambda^2
+\frac{8G_{0}}{g}(
g^3(\frac{1}{\lambda^2}-1)-2g^2
\nonumber \\*
&&+g(\frac{1}{4}\lambda^2-1)+\frac{3}{8}\lambda^2)
+\frac{4G^2_{0}}{g^2}(g^4(1-\frac{2}{\lambda^2})
+g^3(\frac{7}{2}-\frac{2}{\lambda^2})
\nonumber \\*
&&+\frac{g^2}{4}(17-\lambda^2)
+g(1-\frac{3}{4}\lambda^2)
-\frac{9}{16}\lambda^2)]\,.
\label{flpz}
\end{eqnarray}
Here we defined $x=2g\cosh{\lambda} +\lambda^2s(\lambda)$ and the
dimensionless total conductances are now given by
\begin{eqnarray}
  G_{+} & = & \frac{g}{2g+4g\tanh(\lambda/2)/\lambda+3}\,,\nonumber\\
  G_{-} & = & \frac{1}{2g+4g\coth(\lambda/2)/\lambda+3}\,,\nonumber\\
  G_{0} & = & \frac{gx}{gx+4g\cosh{\lambda}+(2g^2+ 3\lambda^2/2)s(\lambda)}\,.\nonumber
\end{eqnarray}

\begin{figure*}
\centerline{\hbox{\epsfxsize=7.in \epsffile{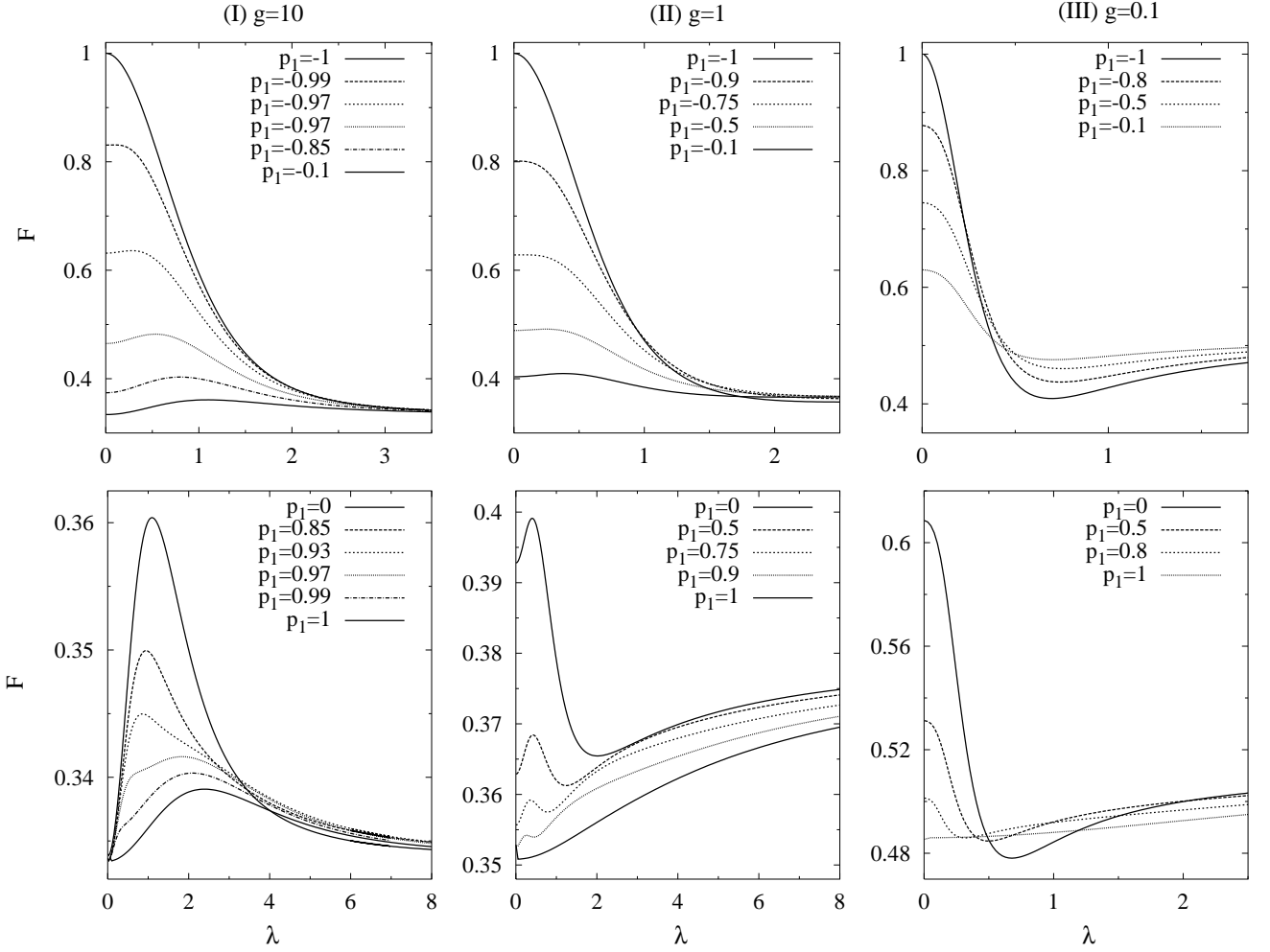} }}
\caption{Fano factor $F$ versus the spin flip scattering intensity
  $\lambda=L/\ell_{\text sf}$ for different polarizations $p_1$ in the
  terminal F$_1$, and perfect polarizations in the other two terminals
  F$_2$, F$_3$: $p_2=p_3=1$. The columns I-III correspond to different
  values of the tunnel contact conductances $g=g_{i}/g_{N}$. Here the
  cross-correlations are related to the Fano factor via
  $S_{23}/|I_{{\text {c}}1}|=(F-1)/4$. For explanation of 
  the various plots, see text.}
\label{zbfig4}
\end{figure*}

The strong dependence of $F$ on $\lambda$, and the magnetic 
configuration is also shown in Fig. \ref{zbfig4}, where $F$ 
versus $\lambda$ is plotted for different polarization $p_1$ 
of the terminal F$_1$, and fixed $p_2=p_3=1$.  
As in Fig. \ref{zbfig2} different columns I-III present results 
for different values of $g$. In each column $p_1$ varies from 
top to bottom in the interval $-1$ to $1$.  In the limit of 
large $\lambda$, the Fano factor tends to the the normal state 
value (\ref{fn}) determined by the conductance $g$ only. 
The deviations from this normal state value at finite 
$\lambda$ depend on $p_1$ and $g$. 
\par
For small values of $g$ (column III), with decreasing $\lambda$ from
large values $F$ first decreases below the normal state value $F_N$ and
then starts to increase again. Thus, there is a minimum of the
Fano factor $F$ occurring at a value of $\lambda$ which continuously
decreases from  $\lambda\sim 1$ to $\lambda=0$ as $p_1$ increases from
$-1$ to $1$. Decreasing $\lambda$ further, $F$ increases to a maximum
value at $\lambda=0$. The maximum of $F$ at $\lambda=0$ continuously
decreases with increasing $p_1$ and becomes a minimum when $p_1=1$.  For
antiparallel fully polarized terminals, i.~e. $p_1=-1$, decreasing
$\lambda$ leads to the strongest variation of $F$, see Eq. (\ref{flpm})
and $F$ reaches the Poissonian value $1$ as $\lambda$ approaches zero.
Comparing plots for $p_1=-1$ of columns (I-III), we see that
this effect is independent of the contact conductances $g$, which is in
agreement with the discussion in connection with Fig.~\ref{zbfig3}.
Increasing $p_1$ from $-1$ decreases the effect of spin flip process in
the current and the noise, and, hence, leads to a reduction of the shot
noise at $\lambda\ll 1$. The maximum value of $F$ thus drops below the
Poissonian value.  
\par 
For large values of $g$ (column I) the maximum of $F$
is shifted from zero to a finite $\lambda \sim 1$ as $p_1$ increases
from $-1$ to $1$, while for small $g$ the maximum $F$ always occurs at
$\lambda=0$ as described above.  For $g\sim1$ (column II) the situation
is in between the large and small $g$ behavior, increasing $\lambda$
leads first to a maximum of the Fano factor followed by a minimum. 
Comparing the first and the second row of Fig.~\ref{zbfig4}, we observe
the effect of spin-flip scattering is more pronounced for negative $p_1$
than for positive $p_1$. This can be understood, because for parallel
magnetization directions the non-equilibrium spin accumulation in the
ferromagnetic wire is smaller. The spin-flip scattering decreases the
spin-accumulation and, hence, has the largest effect for anti-parallel
magnetizations.

\subsection{Cross-correlations}

Let us now discuss the effect of spin flip scattering on the cross
correlations measured between the currents through the terminals F$_2$
and F$_3$. We distinguish two different cases of parallel ($p_2=p_3$)
and antiparallel ($p_2=-p_3$) magnetizations of F$_2$ and F$_3$. For the
parallel case and when $g_{2}=g_{3}$ the two terminals F$_2$, F$_3$ are
completely equivalent and hence $\Delta I_2=\Delta I_3$. This can be
used with Eq. (\ref{sumdeli}) to obtain that in this case the Fano
factor and the cross-correlation factor 
$F_{23}=S_{23}/|2eI_{{\text {c}}1}|$ are
related as (see also Ref.~\cite{cottet04b})
\begin{eqnarray}
  F_{23}=\frac{F-1}{4}.
  \label{ff23}
\end{eqnarray}
Thus $F_{23}$ has the same qualitative dependence on $\lambda$ as $F$.
Since $F \leq 1$ the cross-correlations are always negative as expected
\cite{buettiker:92}. The Fano factor for the perfectly polarized
parallel case is presented in Fig.~\ref{zbfig4} and the
cross-correlations for this case can be deduced from these plots using
Eq.~(\ref{ff23}). We will now analyze the cross-correlations $F_{23}$
for $p_2=p_3=1$ and different values of $p_1$. From Eqs.~(\ref{sijf})
and (\ref{i1m}) one can see that in the limit of large spin-flip
scattering $\lambda \gg 1$, the cross-correlations 
reduces to its all-normal system value
\begin{eqnarray}
F_{23}=-\frac{1}{9}\frac{(4/3)g^3+6g^2+9g+3}{(8/9)g^3+4g^2+6g+3}\,,
\label{f23n}
\end{eqnarray}
which is independent of the polarizations. Alternatively, this result 
could have been obtained using Eqs. (\ref{fn}) 
and (\ref{ff23}).  On lowering $\lambda$ 
the amplitude of the cross-correlations $|F_{23}|$ decreases 
(large $g$, column I) or increases (small $g$, column III) 
with respect to the normal value. 
For $p_1=-1$, $F_{23}$ vanishes in the limit $\lambda \rightarrow
0$, irrespective of the the value of the contacts conductance $g$. 
This case corresponds to the vanishing of the total mean current 
${\bar I}_1$. From this observation we conclude that in the expression 
of $F_{23}$, the cross-correlation $S_{23}$ vanishes faster than 
the mean current ${\bar I}_1$, as $\lambda \rightarrow 0$.
Thus, for $p_1=-1$ $|F_{23}|$ has a minimum at $\lambda=0$ for 
both cases of small and large $g$. Increasing $p_1$ from $-1$, 
the minimum is shifted to a finite  $\lambda \sim 1$ for large 
$g$, it stays always at $\lambda=0$ for small $g$. For small $g$, 
$|F_{23}|$ has also a maximum at $\lambda \sim 1$ which corresponds 
to the minimum of the Fano factor. 
\par   
For $g \sim 1$ (column II) the behavior of $F_{23}$ is between that for small 
and large $g$. Comparing of the plots in top ($p_1<0$) and down 
($p_1>0$) rows in Fig. \ref{zbfig4} shows that the most strong 
variation of the cross-correlation happens when the magnetization 
of F$_1$ is anti-parallel to those of F$_{2,3}$.
\par
\begin{figure*}
\centerline{\hbox{\epsfxsize=7.in \epsffile{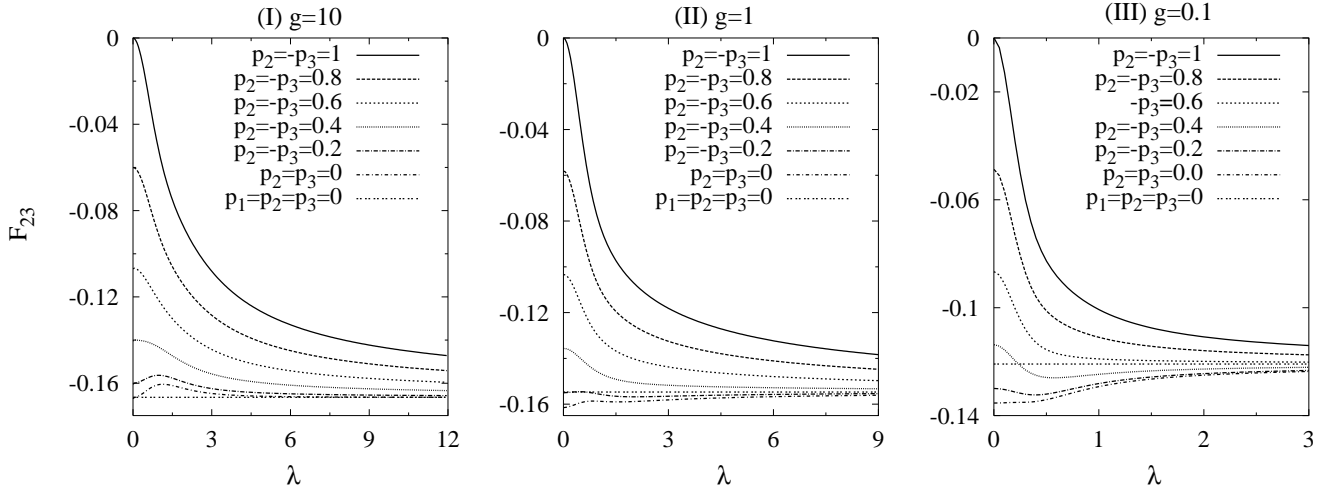} }}
\caption{Cross-correlations of the current measured between
terminals F$_2$ and F$_3$ for anti-parallel configuration of the
polarizations $p_2=-p_{3}$ and for different contact conductances
$g=g_{i}/g_{N}$. The polarization of F$_1$, $p_1=1$ and the
magnitude of the magnetizations $|p_2|=|p_3|$ vary in each 
column.}
\label{zbfig5}
\end{figure*}

For the antiparallel case $p_2=-p_3$ the effect of spin-flip
scattering is more interesting, since it contains the
correlations between currents of opposite spin directions 
produced by the spin-flip scattering. In  Fig. \ref{zbfig5}
we plot $F_{23}$ versus $\lambda$ for different values of
the magnitude $|p_2|=|p_3|=p$ and $g$. We take $p_1=1$ which
corresponds to a maximum spin accumulation in the N-wire 
due to the terminal F$_1$. For $p=1$ the cross-correlations are solely
due to the spin flip-scattering. In this case $F_{23}$ vanishes 
in the limit of $\lambda \rightarrow 0$. 

At finite $\lambda$ the spin-flip scattering induces correlations
between the electrons with opposite spins and hence $F_{23}$ becomes
finite. With increasing $\lambda$ the magnitude of $F_{23}$ increases and
approaches the all-normal system value of $p_1=p_2=p_3=0$ when $\lambda
\gg 1$.  For $p<1$ the value of $|F_{23}|$ for vanishing $\lambda$
depends on the values of $p$ and $g$ as
\begin{eqnarray}
F_{23}=-\frac{9}{8}(1-p^2)\frac{(4/3)g^3+4g^2+4g+1}{(g+1)^3}.
\label{f230l}
\end{eqnarray}
Decreasing $p$ from $1$ to $0$, $|F_{23}|$ increases from zero to a 
maximum value. The maximum absolute value is equal to the normal value for 
large $g$ (column I), while it is larger than the normal value 
for small $g$ (column III). 

\section{Conclusions}

In conclusion, we have investigated the influence of spin polarization
and spin-flip scattering on current fluctuations in a three-terminal
spin-valve system.  Based on a spin-dependent Boltzmann-Langevin
formalism, which accounts for spin-flip scattering in addition to the
usual scattering at impurities and tunnel junctions, we have developed a
semiclassical theory of current fluctuations in diffusive spin-valves.
This theory allows the calculations of spin-polarized mean currents and
correlations of the corresponding fluctuations in multi-terminal systems
of diffusive wires and tunnel contacts. 

We have applied this formalism to a three-terminal system consisting of
a diffusive normal wire connected at the ends to one and two
ferromagnetic terminals, respectively. We have found a strong deviation
of the current correlations from the all-normal system values.  The shot
noise of the total current through the system and the cross-correlations
between currents of two different terminals depend strongly on the
spin-flip scattering rate and the spin polarization and change
drastically with reversing the polarizations in one or more of the
terminals.

The strongest variation of the shot noise occurs, when the polarizations
of the two terminals connected to one end of the normal wire are
antiparallel with respect to the terminal on the other
end. For small spin-flip scattering intensity, the Fano
factor deviates substantially from the normal value and can reach the
full Poissonian value for perfectly polarized terminals even if the
tunnel contact resistances are negligible.  

We have further demonstrated the effect of spin-polarization and
spin-flip scattering on the cross-correlations measured between currents
of two adjacent terminals in two cases where the terminals have parallel
and antiparallel polarizations. For antiparallel orientations of the
contact polarizations, the noise allows a direct determination of
the spin-flip scattering processes in the normal wire. The study of
noise and cross-correlations therefore allows to extract information on
the spin-flip scattering strength, which is of importance for spintronics
applications.

\begin{acknowledgments}

We acknowledge discussions with C. Bruder, T. Kontos and 
H. Schomerus. M.~Z. thanks C. Bruder and his group members 
for hospitality at the University of Basel where this work 
was initiated. W.~B. was financially supported by the RTN 
Spintronics, by the Swiss NSF and the NCCR Nanoscience. 

\end{acknowledgments}

\end{document}